# Wikipedia Vandalism Detection Through Machine Learning: Feature Review and New Proposals*
## Lab Report for PAN at CLEF 2010

Santiago M. Mola Velasco

Private
santiago.mola@bitsnbrains.net

**Abstract** Wikipedia is an online encyclopedia that anyone can edit. In this open model, some people edits with the intent of harming the integrity of Wikipedia. This is known as vandalism. We extend the framework presented in (Potthast, Stein, and Gerling, 2008) for Wikipedia vandalism detection. In this approach, several vandalism indicating features are extracted from edits in a vandalism corpus and are fed to a supervised learning algorithm. The best performing classifiers were LogitBoost and Random Forest. Our classifier, a Random Forest, obtained an AUC of 0.92236, ranking in the first place of the PAN'10 Wikipedia vandalism detection task.

## 1 Introduction

Wikipedia is an online encyclopedia built upon the collaborations of thousands of editors. Its collaboration model is simple: anyone can edit any article at any time. This has made possible the great success of Wikipedia, but it comes with its own problems, one of them being destructive edits. There are many ways in which an edit can be destructive for Wikipedia, such as lobbying, spam, vandalism, tests, etc.

In PAN 2010 Lab's Task 2 we are focused on automatic detection of vandalism. The English Wikipedia defines vandalism as:

> [...] any addition, removal, or change of content made in a *deliberate* attempt to compromise the integrity of Wikipedia. [...] Common types of vandalism are the addition of obscenities or crude humor, page blanking, and the insertion of nonsense into articles.
> Any good-faith effort to improve the encyclopedia, even if misguided or ill-considered, is not vandalism. Even harmful edits that are not explicitly made in bad faith are not vandalism.
> (Wikipedia contributors, 2010a)

The Wikipedia community maintains a project for studying vandalism (Wikipedia contributors, 2010b) and develops bots for automatic detection and reversion. Currently, the most prominent active bot in the English Wikipedia is ClueBot, a rule-based system. (Carter, 2010)

---

*Thanks to the PAN'10 organizers, to the reviewers, to Dmitry Chichkov for his contributions to the PAN'10 group and to Sandra García Blasco for her support and ideas.

ClueBot's rules rely on length increment of the edit and a scoring system based on regular expressions checking vulgarisms, grammar, etc. This system achieves a precision near to 1 but a very low recall. Similar systems can be found in other Wikipedias, such as AVBOT (Rodríguez Posada, 2010) in the Spanish edition.

(Potthast, Stein, and Gerling, 2008) defined Wikipedia vandalism detection as a classification task, proposed a set of discriminating features and provided evaluation of them based on machine learning. Our work is founded in that definition, extending the set of features proposed in their work. We have built a system trying to reproduce previous results, and afterwards, extended it with a new range of features and tuning of previous ones.

We have used the PAN-WVC-10 corpus (Potthast, 2010). The training set of this corpus comprises 15000 annotated edits, 924 of which are labeled as vandalism. Our goal is building a vandalism classifier using this corpus.

The rest of the paper is structured as follows. In Section 2 of this article, we will describe the preprocessing we have applied to the data. In Section 3 we discuss all edit features that we have used. In Section 4 we present the classifiers we have used. In Section 5 we will evaluate our features and classifiers and present the classifier we have used for the PAN'10 result submission. Finally, we will present our conclusions in Section 6.

## 2 Preprocessing

In our feature framework, we have defined different representations of edits to use in different features. In this section we will describe these representations and the steps to construct them.

First, we defined a tokenization function. A token can be a word, a number, a punctuation mark or a wiki-syntax element. When describing features, we will consider a *word* as any of these tokens. The following symbols are considered independent tokens: ., ,, :, ;, ", «, », ', |, ?, !, =, (, ), *, [[, ]], [, ], {{, }}, { and }.

We also need to produce diffs between the old and new revisions of an edit. To that end, we used google-diff-match-patch.[1]

Using these funcions, we define the following representations of edit text:

- **Old text** and **new text**. The old and new without any preprocess.
- **Case-sensitive inserted words**. The set of inserted words.
- **Inserted words**. The set of inserted words, and converted to lowercase.
- **Concatenated inserted words**. All inserted words concatenated and separated with spaces. This is defined both case-sensitive and insensitive.
- **Inserted text**. Inserted lines as reported by the diff algorithm.

---

[1] Available at `google-diff-match-patch`. Behaviour of different parameters of the diff function has not been studied yet. The parameters used were `Patch_Margin` set to 0, `Diff_EditCost` to 6 and `Match_Distance` to 500.

## 3 Edit Features

In this section, we describe our set of features. Features based on dictionaries will be explained separately. Features marked with * were already defined in (Potthast, Stein, and Gerling, 2008) and those marked with † are modifications of features also defined in that work.

**Anonymous*** Wether the editor is anonymous or not.
>Vandals are likely to be anonymous. This feature is used in a way or another in most antivandalism working bots such as ClueBot and AVBOT. In the PAN-WVC-10 training set (Potthast, 2010) anonymous edits represent 29% of the regular edits and 87% of vandalism edits.

**Comment length*** Length in characters of the edit summary.
>Long comments might indicate regular editing and short or blank ones might suggest vandalism, however, this feature is quite weak, since leaving an empty comment in regular editing is a common practice.

**Upper to lower ratio†** Uppercase to lowercase letters ratio, i.e., $\frac{1+|upper|}{1+|lower|}$.
>Vandals often do not follow capitalization rules, writing everything in lowercase or in uppercase.

**Upper to all ratio†** Uppercase letters to all letters to ratio, i.e., $\frac{1+|upper|}{1+|lower|+|upper|}$.

**Digit ratio** Digit to all characters ratio, i.e., $\frac{1+|digit|}{1+|all|}$.
>This feature helps to spot minor edits that only change numbers, which might help to find some cases of subtle vandalism where the vandal changes arbitrarily a date or a number to introduce misinformation.

**Non-alphanumeric ratio** Non-alphanumeric to all characters ratio, i.e., $\frac{1+|nonalphanumeric|}{1+|all|}$.
>An excess of non-alphanumeric characters in short texts might indicate excessive use of exclamation marks or emoticons.

**Character diversity** Measure of different characters compared to the length of inserted text, given by the expression $length^{\frac{1}{different\,chars}}$.
>This feature helps to spot random keyboard hits and other non-sense. It should take into account QWERTY keyboard layout in the future.

**Character distribution†** Kullback-Leibler divergence of the character distribution of the inserted text with respect the expectation. Useful to detect non-sense.

**Compressibility†** Compression rate of inserted text using the LZW algorithm.[2]
>Useful to detect non-sense, repetitions of the same character or words, etc.

**Size increment** Absolute increment of size, i.e., $||new|-|old||$.
>The value of this feature is already well-established. ClueBot uses various thresholds of size increment for its heuristics, e.g., a big size decrement is considered an indicator of blanking.

**Size ratio*** Size of the new revision relative to the old revision, i.e., $\frac{1+|new|}{1+|old|}$.
>Complements size increment.

---

[2] LZW was chosen after evaluating the behaviour of LZW, gzip and bzip2, although, an exhaustive comparison is still pending. We used the TIFF LZW algorithm (Adobe Developers Association, 1992) as implemented in python-lzw 0.01 by Joe Bowers, available at `http://www.joe-bowers.com/static/lzw/`.

**Average term frequency*** Average relative frequency of inserted words in the new revision.

In long and well-established articles too many words that do not appear in the rest of the article indicates that the edit might be including non-sense or non-related content.

**Longest word*** Length of the longest word in inserted text.

Useful to detect non-sense.

**Longest character sequence*** Longest consecutive sequence of the same character in the inserted text.

Long sequences of the same character are frequent in vandalism (e.g. *aaggggghhh-hhhh!!!!!*, *soooooo huge*).

Size increment and ratio are closely related to all features, as can be seen in ClueBot and AVBOT source code, where thresholds for these measures are coded deep into most heuristics. This should be considered when building a classifier.

Upper to all, upper to lower, digit and non-alphanumeric ratios were computed using case-sensitive concatenated inserted words. These could also be computed using a diff or computing the difference of their values in the old and new text. Each method produces different results. We chose only inserted words for efficiency.

Antivandalism bots use lists of regular expressions with weights in order to compute a vandalism score. In (Potthast, Stein, and Gerling, 2008) a similar approach is described, where vulgarism frequency and impact are considered. Vulgarism frequency is the frequency of vulgarisms in an edit text relative to all words in the edit. Vulgarism impact is the percentage by which an edit increases the number of vulgarisms in the text. In the same way, pronoun frequency and impact were defined.

We have defined features analogous to vulgarism frequency and vulgarism impact, for different categories of words:

**Vulgarisms** Vulgar and offensive words, e.g., *fuck*, *suck*, *stupid*.

**Pronouns** First and second person pronouns, including slang spellings, e.g., *I*, *you*, *ya*.

**Biased** Colloquial words with high bias, e.g., *coolest*, *huge*.

**Sex** Non-vulgar sex-related words, e.g., *sex*, *penis*, *nipple*.

**Bad** Hodgepodge category for colloquial contractions (e.g. *wanna*, *gotcha*), typos (e.g. *dosent*), etc.

**All** A meta-category, containing vulgarisms, pronouns, biased, sex-related and bad words.

**Good** Words rarely used by vandals, mainly wiki-syntax elements (e.g. *__TOC__*, *<ref>*)

Word counters were computed using the inserted words in lowercase. In future work, it would be desirable to consider caseness and style. For example, *dick* is very likely to be slang for penis and *DICK* is a strong indicator of vandalism, but *Dick* is more likely to be the diminutive for Richard or a surname. Our current system does not make this distinction.

# 4 Classification

Classification has been conducted using the Weka framework (Hall et al., 2009). After preliminary evaluations, we have tried to choose classifiers which fullfil all or most of the following conditions: *a)* require little or no preprocessing of data, *b)* require little parameter adjustment, *c)* do implicit feature selection, *d)* are resistant to noise and outliers and *e)* are resistant to severe class imbalance.

Our baseline classifier is C4.5 decision tree (Quinlan, 1993) which is a well-established algorithm and, to some extent, fullfils our criteria. LogitBoost (Friedman, Hastie, and Tibshirani, 2000) and Random Forest (Breiman, 2001) are attractive because of their implicit feature selection, generalization properties and a low number of parameters. We expect Random Forest to properly exploit relations between features such as those described in Section 3. Random Forest has another interesting property: it uses an internal test error estimation known as out of bag error which makes cross-validation prescindible.

We also chose Support Vector Machines (SVM) because of its resistance to class imbalance (Japkowicz and Stephen, 2002), however, SVM has been discarded for inclusion in the PAN'10 results because of its high computational demands and amount of parameter to adjust.

# 5 Evaluation

The following measures have been evaluated in every experiment: Precision, Recall, F-Measure and Area Under ROC Curve (AUC).

In Table 1 we can see the performance of a C4.5 decision tree built for each feature. The C4.5 algorithm, considering only one feature and with severe class imbalance, will often produce a majority class classifier. In order to prevent this, we changed the weight of vandalism edits so the cost of a misclassified vandalism edit to be 10 times higher than a misclassified regular edit. This is meant to be orientative, not an accurate ranking of features.

In Table 2 we can see the performance of all the classifiers considering all features. Every classifier has been evaluated with 10-fold cross-validation. Parameters used for classifiers were: [3]

**C4.5** minimum number of instances per leaf is 6, Laplace smoothing is used for predicted probabilites.
**LogitBoost** 100, 500 and 1000 boost iterations. The weak classifier is a Decision Stump.
**Random Forest** 100, 500 and 1000 iterations.

LogitBoost and Random Forest perform similarly in this task. Random Forest improves F-Measure and AUC as we increase the number of iterations. While LogitBoost shows slightly better performance than Random Forest, its results are less stable and decrease while increasing number iterations, suggesting a problem of overfitting. This is, presumably, because of LogitBoost's sensitiveness to noise and outliers.

---

[3]Any omitted parameter uses default parameters in Weka. For more information, check Weka API documentation at http://weka.sourceforge.net/doc.stable/.

**Table 1.** Evaluation of individual features with a C4.5 decision tree.

| Feature | Precision | Recall | F-Measure | AUC |
|---|---|---|---|---|
| Anonymous | 0.166 | 0.872 | 0.278 | 0.78 |
| Comment length | 0.102 | 0.663 | 0.177 | 0.661 |
| Upper to lower ratio | 0.131 | 0.629 | 0.217 | 0.755 |
| Upper to all ratio | 0.129 | 0.621 | 0.214 | 0.731 |
| Digit ratio | 0.128 | 0.718 | 0.217 | 0.75 |
| Non-alphanumeric ratio | 0.117 | 0.629 | 0.197 | 0.729 |
| Character diversity | 0.121 | 0.689 | 0.206 | 0.741 |
| Character distribution | 0.11 | 0.698 | 0.19 | 0.71 |
| Compressibility | 0.095 | 0.29 | 0.143 | 0.627 |
| Size increment | 0.111 | 0.669 | 0.191 | 0.69 |
| Size ratio | 0.1 | 0.195 | 0.132 | 0.572 |
| Average term frequency | 0.207 | 0.021 | 0.037 | 0.503 |
| Longest word | 0.124 | 0.712 | 0.211 | 0.742 |
| Longest character seq. | 0.115 | 0.514 | 0.189 | 0.628 |
| All frequency | 0.762 | 0.353 | 0.482 | 0.661 |
| All impact | 0.501 | 0.377 | 0.43 | 0.662 |
| Vulgarism frequency | 0.904 | 0.214 | 0.346 | 0.597 |
| Vulgarism impact | 0.618 | 0.227 | 0.332 | 0.599 |
| Bad word frequency | 0.747 | 0.08 | 0.145 | 0.532 |
| Bad word impact | 0.348 | 0.111 | 0.169 | 0.539 |
| Biased frequency | 0.623 | 0.082 | 0.145 | 0.536 |
| Biased impact | 0.311 | 0.173 | 0.223 | 0.561 |
| Sex frequency | 0.717 | 0.041 | 0.078 | 0.517 |
| Sex impact | 0.537 | 0.048 | 0.087 | 0.518 |
| Pronoun frequency | 0.588 | 0.152 | 0.241 | 0.562 |
| Pronoun impact | 0.311 | 0.173 | 0.223 | 0.561 |
| Goodword frequency | 0 | 0 | 0 | 0.545 |
| Goodword impact | 0.149 | 0.106 | 0.124 | 0.584 |

**Table 2.** Classifier evaluation.

| Classifier | Precision | Recall | F-Measure | AUC |
|---|---|---|---|---|
| C4.5 | 0.773 | 0.543 | 0.638 | 0.93 |
| Random Forest 100 it. | 0.855 | 0.568 | 0.683 | 0.957 |
| Random Forest 500 it. | 0.86 | 0.565 | 0.682 | 0.962 |
| Random Forest 1000 it. | 0.861 | 0.568 | **0.684** | **0.963** |
| LogitBoost 100 it. | 0.845 | 0.566 | 0.678 | **0.966** |
| LogitBoost 500 it. | 0.82 | 0.606 | **0.697** | 0.964 |
| LogitBoost 1000 it. | 0.795 | 0.611 | 0.691 | 0.961 |

Our final classifier was a Random Forest of 1000 trees, each one built while considering 5 random features[4], obtaining an AUC of 0.92236 when evaluated with the PAN-WVC-10 testing corpus.

---
[4] That is, $\log_2 M + 1$, where $M$ is the number of features.

## 6  Conclusions

In this work, we revisited (Potthast, Stein, and Gerling, 2008) approach on Wikipedia vandalism detection, tuning some features and introducing others. We further explored features based on word lists, expanding them beyond vulgarisms and improving results by creating new categories. We performed a brief comparison of different supervised learning techniques for this task. Random Forest and LogitBoost produced good results.

Our final classifier, a Random Forest, obtained an AUC of 0.92236, which was ranked in the first place in PAN'10 for the vandalism detection task. Precision and recall on the training test were 0.861 and 0.568, respectively. This is far better than current antivandalism systems in terms of F-Measure, but it has neither precision near to 1, which is required for a bot in order to operate autonomously, nor a recall high enough to be used as a filtering system for human editors. In any case, our results suggest that this machine learning based approach is a suitable path for the next-generation antivandalism bots.

Future work might include a more exhaustive feature evaluation; building bigger word lists using free lexical resources such as Wiktionary and WordNet and infer their weights; introducing a feature for detecting random keyboard hits considering QWERTY keyboard layout; introducing heuristics from current antivandalism bots and experiment with measures based on ngram language models.